\documentclass[preprint showpacs,preprintnumbers,amsmath,amssymb]{revtex4}
\usepackage{xcolor}
\usepackage{graphicx}% Include figure files
\usepackage{dcolumn}% Align table columns on decimal point
\usepackage{bm}% bold math
\usepackage{multirow}
\usepackage{natbib}
\usepackage{color}
\usepackage{cases} 

\begin{document}
\preprint{APS/123-QED}

%\title{ Optical shielding of ultracold K-Cs binary collision}
\title{Dissociative recombination of N$_2$H$^+$: A revisited study}

\author{J. Zs Mezei$^{1,2}$}
\email[]{mezei.zsolt@atomki.hu}
\author{M. A. Ayouz$^{3}$}
\author{A. Orb\'an$^{1}$}
\author{A. Abdoulanziz$^{2}$}
\author{D. Talbi$^{4}$}
\author{D. O. Kashinski$^{5}$}
\author{E. Bron$^{6}$}
\author{V. Kokoouline$^{7}$}
\author{I. F. Schneider$^{2,8}$}
\affiliation{$^{1}$Institute for Nuclear Research (ATOMKI), H-4001 Debrecen, Hungary}%
\affiliation{$^{2}$LOMC CNRS-UMR6294, Universit\'e le Havre Normandie, F-76058 Le Havre, France}%
\affiliation{$^{3}$LGPM CNRS-EA4038, Universit\'e Paris-Saclay - CentraleSupelec, 91190 Gif-sur-Yvette, France}%
\affiliation{$^{4}$LUPM CNRS-UMR5299, Universit\'e de Montpellier, 34095 Montpellier, France}%
\affiliation{$^{5}$Photonics Research Center, United States Military Academy, 10996 West Point, New York, USA}%
\affiliation{$^{6}$LERMA CNRS-UMR8112, Observatoire de Paris - Universit\'e PSL, 92195 Meudon, France}%
\affiliation{$^{7}$Dept. of Physics, University of Central Florida, 32816 Orlando, Florida, USA}%
\affiliation{$^{8}$LAC CNRS-UMR9025, Universit\'e Paris-Saclay, F-91405 Orsay, France}%
\date{\today}

\begin{abstract}
Dissociative recombination of N$_2$H$^+$ is explored in a two-step theoretical study. In a first step, a diatomic (1D) rough model with frozen NN bond and frozen angles is adopted, in the framework of the multichannel quantum defect theory (MQDT). The importance of the indirect mechanism and of the bending mode is revealed, in spite of the disagreement between our cross section and the experimental one. In a second step, we use our recently elaborated 3D approach based on the normal mode approximation combined with R-matrix theory and MQDT. This approach results in satisfactory agreement with storage-ring measurements, significantly better at very low energy than the former calculations. 
\end{abstract}

%\pacs{33.80. -b, 42.50. Hz}% PACS, the Physics and Astronomy
                             % Classification Scheme.
%\keywords{coupled-channel, optical shielding, KCs}%Use showkeys class option if keyword

                              %display desired
\maketitle

\section{Introduction}\label{sec:intro}
 Diazenylium, or N$_2$H$^+$, a three-atomic inorganic cation, is one of the first charged species observed in the interstellar medium (ISM), and - in a smaller percentage - in terrestrial environments. It gives astronomers information about the fractional ionization of interstellar molecular clouds and about the chemistry therein, and it is often used as a tracer for molecules that are not as easily detected due to the lack of permanent dipole moment, such as N$_2$~\cite{caselli1995}. 
 
 N$_2$H$^+$ has been detected in a variety of interstellar environments including dark clouds~\cite{turner1974}, translucent clouds~\cite{turner1995}, protostellar cores~\cite{caselli2002}, and photodissociation regions~\cite{fuente1993}. 
 In the ISM N$_2$H$^+$ is mainly produced in proton transfer reaction to N$_2$ from H$_3^+$ and can be lost via proton transfer to abundant molecules like CO or recombining with slow electrons. 
 N$_2$H$^+$ plays also an important role in the nitrogen-rich planetary atmospheres of solar planets and/or moons and exoplanets. For example, in the N$_2$-dominated ionosphere of Titan, N$_2$H$^+$ is formed from the reaction between N$^+_2$ and H$_2$, but is quickly lost via in electron recombination or proton transfer to CH$_4$~\cite{vuitton2007}. To predict the abundance of N$_2$ based on N$_2$H$^+$ measurements it is necessary to have an insight into the production and destruction mechanisms of N$_2$H$^+$. 
 
 In the present paper we  plan to study the dissociative recombination (DR) of N$_2$H$^+$ in collisions with slow electrons. This process may follow two paths as follows:
\begin{equation}
\label{eq:reacDR}
\mbox{N}_2\mbox{H}^{+} + e^{-} \longrightarrow\begin{cases}
\mbox{N}_2 + \mbox{H} & \\
\mbox{NH}  + \mbox{N} & \\
\end{cases}
\end{equation}

The DR of N$_2$H$^+$ has been studied experimentally quite intensively. A Flowing Afterglow Langmuir Probe (FALP) experiment~\cite{smith1984} resulted in  a thermal rate coefficients of the reaction equal to $1.7\times10^{-7}$ cm$^3\cdot$s$^{-1}$ at room temperature. This was somewhat confirmed by later FALP measurements~\cite{poterya2005,lawson2011}, providing a nearly constant rate coefficient of about $2.8 \times 10^{-7}$ cm$^3\cdot$s$^{-1}$ over the temperature range from 200 to 500 K. Moreover, by using absorption technics for the final products \cite{adams1991}, it was found that the DR of N$_2$H$^+$  predominantly goes into the $\mbox{N}_2+\mbox{H}$ branch of eq.~(\ref{eq:reacDR}). In contrast to these findings, in an experiment performed on CRYRING~\cite{geppert2004},  $64\%$ of the DR  was reported in the $\mbox{NH}+\mbox{N}$ branch of eq.~(\ref{eq:reacDR}). 
 This was  disproved eventually by a new flowing afterglow measurement~\cite{molek2007} and by a second CRYRING measurement~\cite{Vigren2012} (providing DR cross sections up to 10 eV collision energies), both confirming that more than $90\%$ of the DR goes into the $\mbox{N}_2+\mbox{H}$ branch. 

In addition to these experimental studies, extensive theoretical effort has been provided for the DR of N$_2$H$^+$. D. Talbi {\it et al}. have published a series of structure calculations~\cite{talbi2007,Hickman2011,Kashinski2012,kashinski2015} performed at linear geometries with the NN/NH bond lengths frozen in the framework of the block diagonalisation method~\cite{pacher1988} combined with GAMESS~\cite{schmidt1993}, identifying and characterising both possible - experimentally explored - DR pathways, and showing that  N$_2$ + H should be favoured over  N + NH because of the absence of a favourable dissociating state for the N$_2$ bond breaking~\cite{talbi2009}. Three repulsive N$_2$H valence states whose potential energy curves (PECs) cross the ion's one far from the favourable Franck-Condon region were found. This led to the expectation of a small {\it direct} DR cross section for vibrationally relaxed $(v^+=0)$ ions.
 Meanwhile, a simple model developed by Pratt and Jungen~\cite{jungen2008} (successfully applied for HCO$^+$) shows that the DR of polyatomic molecular cations may be quite efficient when the {\it indirect} pathway only is relevant. 

A more comprehensive alternative method has been elaborated by Douguet {\it et al.}~\cite{douguet2011} and applied to  N$_2$H$^+$\cite{Santos2014}, but also to other 
polyatomic cations - H$_3$O$^+$, H$_3^+$~\cite{douguet2011,douguet2012}, HCO$^+$ and  CH$_3^+$\cite{douguet2012} - providing cross sections and rate coefficients in good agreement with
experimental data over a wide energy range. More recently, this method has been used in order to produce rate coefficients for species of interest for cold plasma applications - BF$_2^+$\cite{slava2018} - and for large molecules of astrochemical interest - CH$_2$NH$_2^+$\cite{yuen2019} and NH$_2$CHOH$^+$\cite{ayouz2019}. It is based on the computation of the scattering matrix using electron scattering calculations combined with vibronic frame transformation and takes into account all vibrational normal modes of the cations. Fonseca dos Santos {\it et al.} in \cite{Santos2014} have presented results for the indirect dissociative recombination of N$_2$H$^+$ using electron scattering calculations with the complex Kohn variational method and they found DR cross sections agreeing well with the CRYRING measurements~\cite{Vigren2012}.

In a later study, Fonseca dos Santos {\it et al.}~\cite{Santos2016} have published results on the direct and indirect mechanisms of dissociative recombination of N$_2$H$^+$. Electron-scattering calculations responsible for the indirect mechanism were performed using the complex Kohn variational method (scattering matrix, energy positions and autoionization widths of resonant states), while the cross section for the direct dissociation along electronic resonant states is computed with wave-packet calculations using the Multi Configuration Time-Dependent Hartree method~\cite{mctdh} with all three internal degrees of freedom considered. The calculated direct DR cross sections are found smaller by more than an order of magnitude compared to those obtained for the indirect mechanism.

The present paper is structured as follows: In the section following this introduction, we describe the use of our method based on the Multichannel Quantum Defect Theory, in the framework of a 1D diatomic model. This approach is based on the molecular structure data obtained by quantum chemistry calculations of Talbi {\it et al.}~\cite{talbi2007,Kashinski2012}, and demonstrates the importance of the indirect mechanism. 

In the next section, we revisit the polyatomic (3D) model presented in \cite{Santos2014}, and perform a new study, within the normal mode approximation, combined with the R-matrix theory and the vibronic frame transformation. 
We present the obtained new DR cross section and compare with the preceding theoretical results and with the storage ring measurements.

The paper ends with conclusions and perspectives.

\section{Dissociative recombination of N$_2$H$^+$: A diatomic 1D model}\label{sec:1D}
In the present section we present our results obtained using the Multichannel Quantum Defect Theory  in the frame work of the 1D or {\it diatomic} model based on molecular data set calculated by Talbi {\it et al.}~\cite{talbi2007,Kashinski2012}.

The efficiency of our stepwise MQDT method in modelling the electron/diatomic cation collisions has been proved in many previous studies on different species, like H$_2^+$ and its isotopologues~\cite{motapon2008,waffeutamo2011,chakrabarti2013,motapon2014,Epee2015,Epee2022}, N$_2^+$~\cite{little2014,abdoulanziz2021}, and many more.  
The general ideas of our approach were already presented in detail earlier, see for example~\cite{mezei2019} and, therefore, here we restrict ourselves to the description of the mechanisms.

The reaction~(\ref{eq:reacDR}) involves \textit{ionization} channels - characterising the scattering of an electron on the target cation - and \textit{dissociation} channels - relating to atom-atom scattering. The mixing of these channels results in quantum interference of the \textit{direct} mechanism - in which the capture takes place into a doubly excited dissociative state of the neutral system - with the \textit{indirect} one - in which the capture occurs via a Rydberg bound state of the molecule belonging to a \textit{closed} channel, this state being predissociated by the dissociative one. 
In both mechanisms the autoionization - based on the existence of \textit{open}  ionization channels - is in competition with the predissociation, and can lead to the excitation or to the de-excitation of the cation.

Depending on the total energy of the system these ionization channels can be {\it open} - either as entrance channels, describing the incident electron colliding the ion in its ground electronic state, or  exit channels, describing the autoionization, i.e. resonant elastic scattering, ro-vibrational excitation and de-excitation  - or {\it closed} - describing the resonant temporary captures into Rydberg states. 

The MQDT treatment of DR requires the {\it a priori} knowledge of the PECs of the ion ground state and of the relevant doubly excited, dissociative states of the neutral molecule, as well as for the Rydberg series of mono-excited states represented by the quantum defects. The driving forces  of the recombination and excitation processes are the electronic couplings that connects the dissociative and ionization continua.

\begin{figure}[t]
\centering
\includegraphics[width=0.8\linewidth]{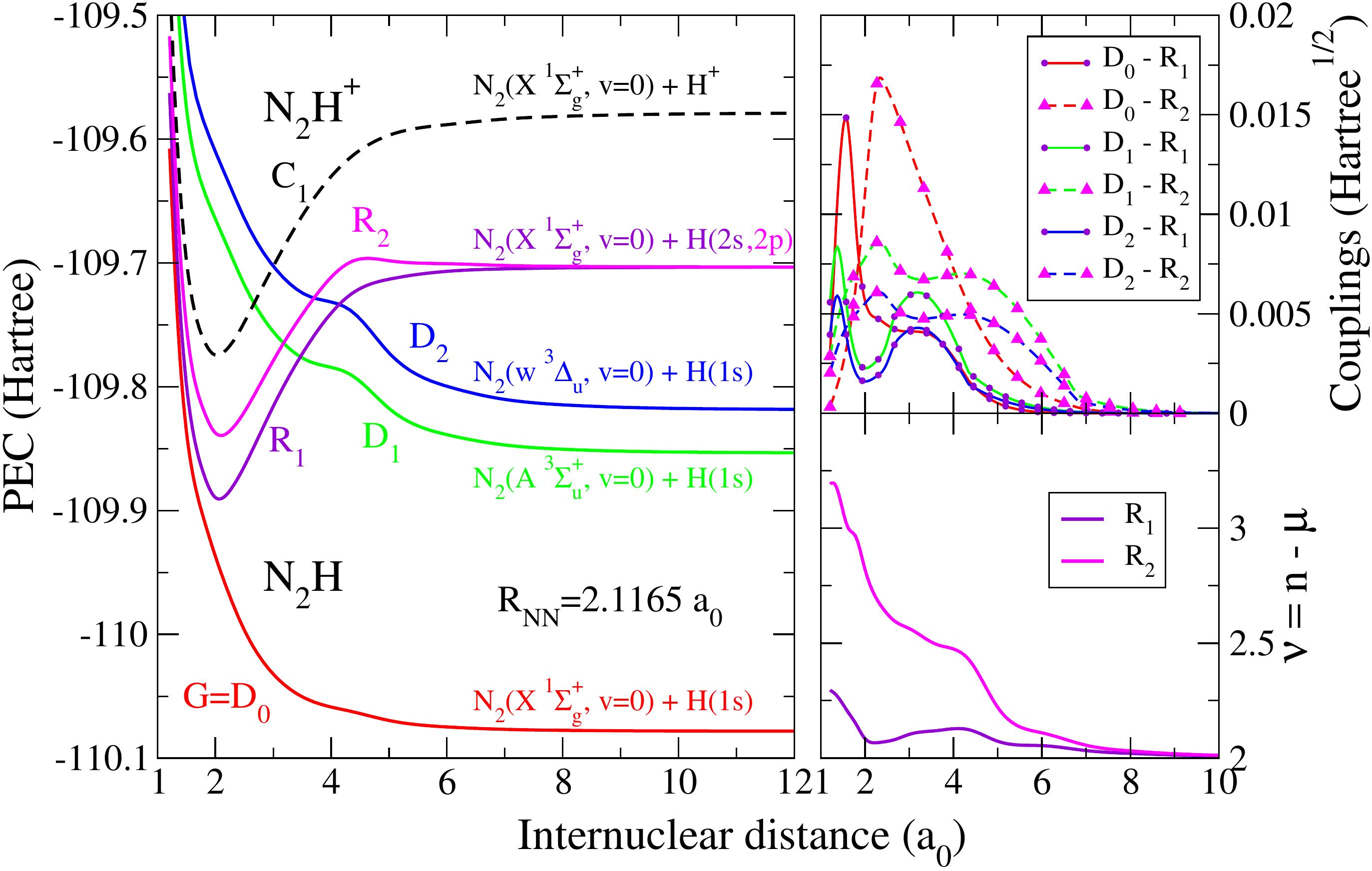}
 \caption{ Molecular data sets used in our 1D - MQDT calculation compiled from~\cite{talbi2007,Kashinski2012}. Left panel: 
 Potential energy curves of the molecular cation and neutral system relevant for DR.  Upper-right and lower-right panels: The Rydberg-valence electronic couplings and quantum defects.
 }
\label{fig:moldata}
\end{figure}

Figure~\ref{fig:moldata} contains all the molecular data relevant for DR via the ,,N$_2$'' pathway (frozen NN bond and frozen angles), the upper branch of eq. (\ref{eq:reacDR}) calculated by Talbi {\it et al.}~\cite{talbi2007,Kashinski2012}. 
On the left in black, we displayed the PEC of the ground electronic state of N$_2$H$^+$ (C$_1$), and in red, green  and blue respectively the PECs of the ground (D$_0$) and two lowest excited repulsive states  of the neutral molecule (D$_1$, D$_2$), having dissociative character.  The violet (R$_1$) and magenta (R$_2$) curves correspond to two representatives ($s$ and $p$ partial waves) of the mono-excited Rydberg series. In the figure we have marked the asymptotic limits for each PEC. In addition to the PECs we present the Rydberg-valence electronic couplings (upper right graph) and the effective principal quantum numbers of the two Rydberg states (lower right graph). They belong to two different Rydberg series, corresponding to different dominant partial waves.

One may notice that, at low collision energies, neither of the three PECs of the dissociative states (D$_0$, D$_1$ or D$_2$) has a favourable crossing with that of the ion. This suggests that the direct DR cross section, at least according to this model, is relatively small. 

Indeed, in the first order and assuming that the two dissociative channels D$_1$ and D$_2$ are independent, the  direct DR cross section via one of theses channels reads as~\cite{giusti1980}:

\begin{figure}[t]
\centering
\includegraphics[width=0.8\linewidth]{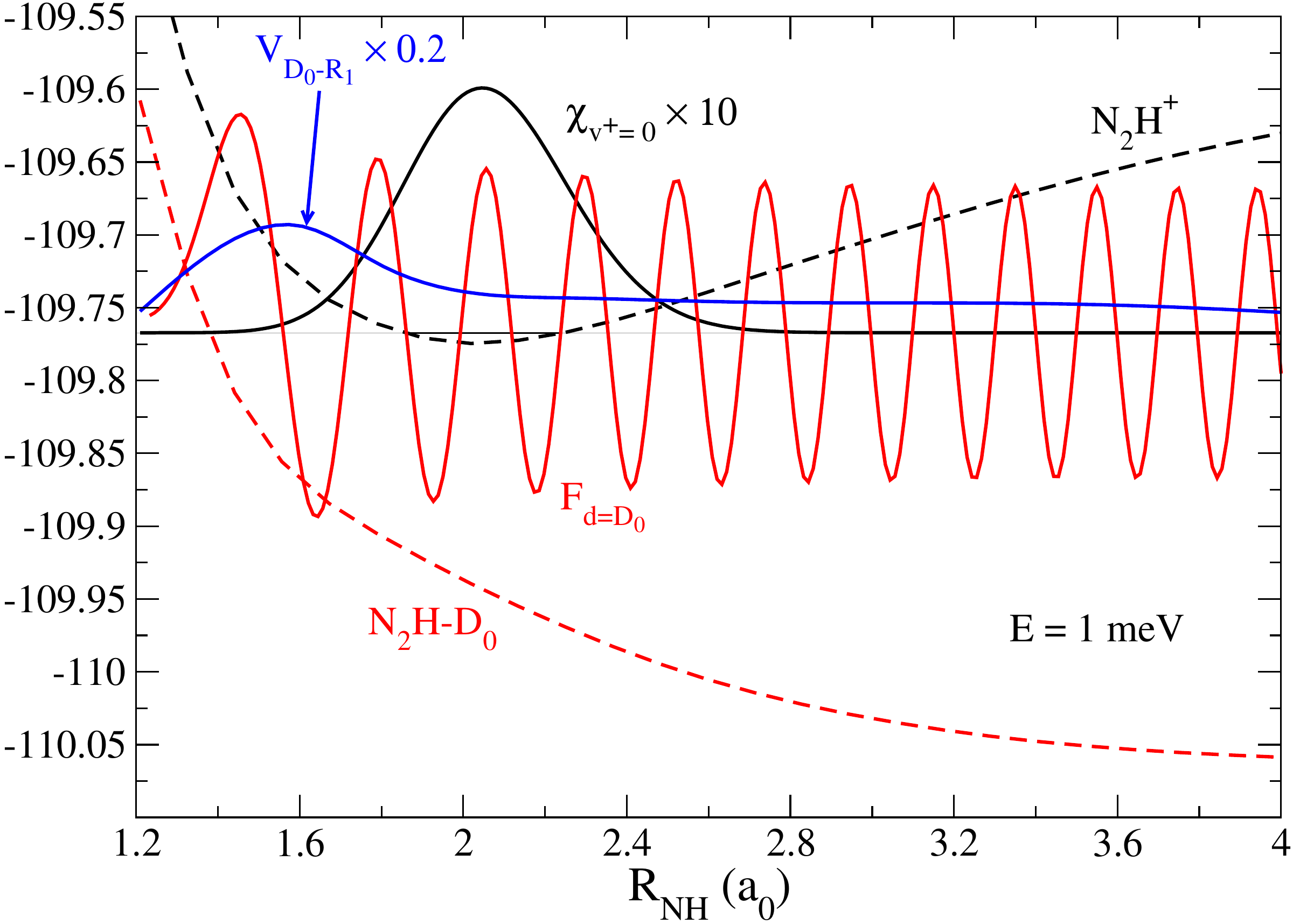}
 \caption{
Main data relevant for the estimation of the  direct DR cross section in first order for 1 meV collision energy, according to eq.~(\ref{eq:dirxs}). The dashed lines stand for the PECs of the ion (black) and for the lowest  dissociative state (red) labeled with D$_0$ in fig.~\ref{fig:moldata}. The black solid line stands for the vibrational wave function of the ion ($\chi_{v^+}$ for $v^+=0$), the blue solid line represents the Rydberg-valence coupling ($V^{el}(R)$, as in the  upper right-hand side graph of fig.~\ref{fig:moldata}), and the red curve stands for the continuum wave function of the dissociative state regular in the origin ($F_d$).
 }
\label{fig:dirxs}
\end{figure}

\begin{equation}\label{eq:dirxs}
\sigma^{direct}_{diss \leftarrow {v^+_{i}}} =\frac{\pi \rho}{4\varepsilon}\frac{4(\xi_{{v}_i^+})^2}{\left[1+\sum_{v^+}\left(\xi_{v^+}^2\right)^2\right]^2},\,\,\mbox{where}\,\,\xi_{v^+}=\pi\langle \chi_{v^+} \left\vert V^{el}(R) \right\vert F_d \rangle.
\end{equation}
\noindent 
Here $\varepsilon$ stands for the collision energy, $v_i^+$ is the initial vibrational level of the target cation, $\rho$ gives the multiplicity ratio between neutral molecule and molecular cation states, $\chi_{v^+}$ and $F_d$ are the vibrational and regular continuum wave functions of the ion ground state and dissociative neutral molecular state ( the later calculated in $\varepsilon$), while $V^{el}(R)$ represents the electronic coupling between the ionization and the dissociation continua. The sum in eq. (\ref{eq:dirxs}) is performed on the vibrational levels corresponding to the open ionization channels.

The elements related to the direct cross section are presented in fig.~\ref{fig:dirxs}. One can see that the overlap between the vibrational wave function of the ground electronic state of the ion, the continuum wave function of the D$_0$ dissociative state of the neutral and its electronic coupling with the ionization continuum will lead to a small numerical value, and thus to a very small direct DR cross section.

\begin{figure}[t]
\centering
\includegraphics[width=0.8\linewidth]{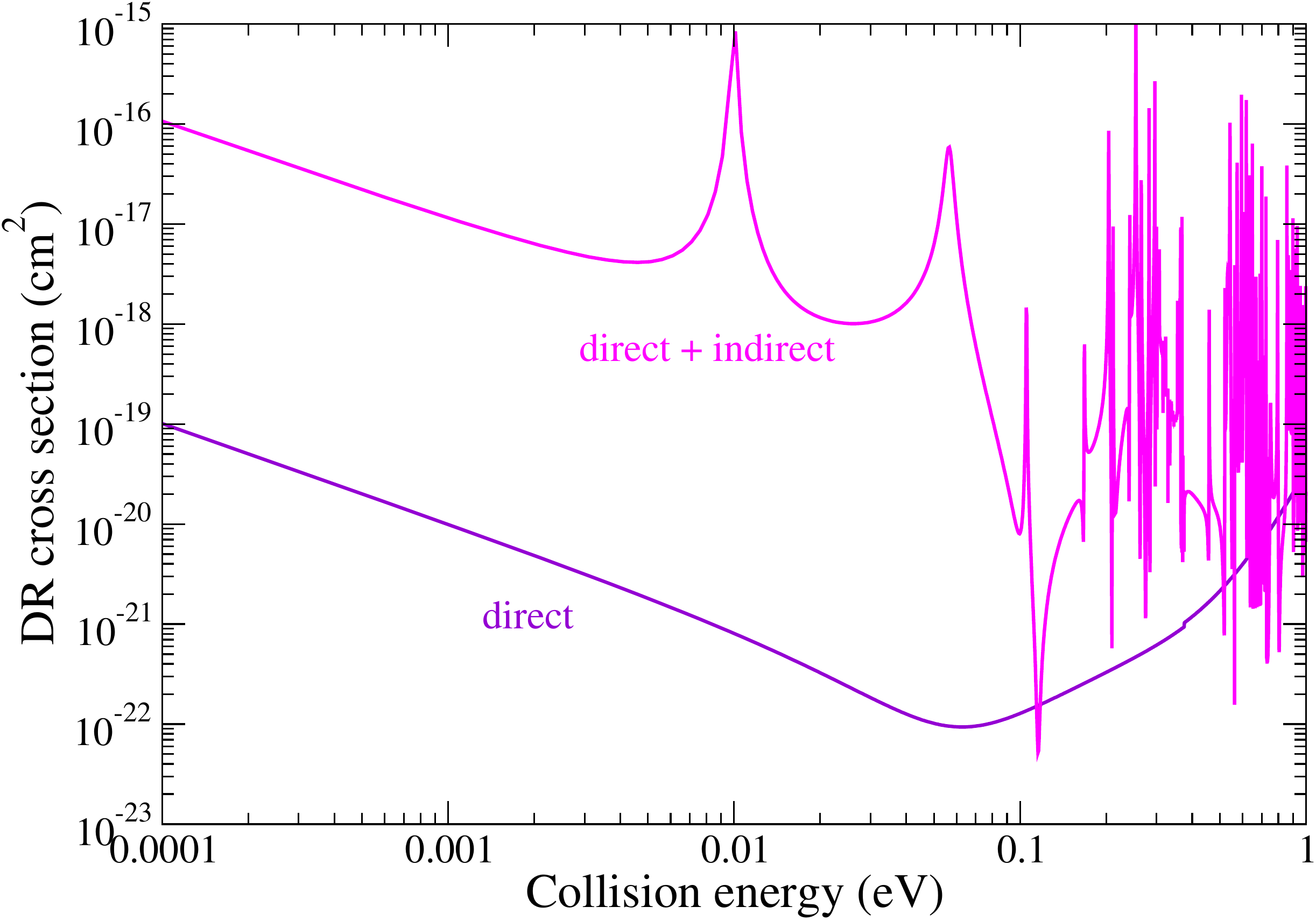}
 \caption{
DR of N$_2$H$^+$ at low energy of the  incident electron within the 1D MQDT model. Violet: direct process. Magenta: total process.
 }
\label{fig:1dxs}
\end{figure}

Nevertheless, the indirect mechanism can be at the origin of the DR process, as it is for example for H$_3^+$~\cite{slava2001} or HeH$^+$~\cite{curic2020}. In order to understand its importance in the case of N$_2$H$^+$, we have applied our stepwise MQDT method using the molecular data set presented in fig.~\ref{fig:moldata}. The total cross section, (the reader is referred for details to~\cite{mezei2019}) is given by:

\begin{equation}\label{eq:drxsec1}
	\sigma_{diss \leftarrow {v^+_{i}}} = \frac{\pi}{4\varepsilon} \sum_{{l,\Lambda}}
	\rho^{{\Lambda}} \sum_{j} \left\vert S^{{\Lambda}}_{d_j, {l}v^+_{i}} \right\vert^2.
\end{equation}

\noindent Here, $l$ stands for the partial wave of the incoming/outgoing electron and $\Lambda$ for the projection of the angular momentum on the molecular axis. The sum is performed over all partial waves and symmetries as well as all energetically open dissociative ($d_j$) states.

Applying our stepwise MQDT method we have calculated the direct and total dissociative recombination (eq. (\ref{eq:drxsec1})) cross sections of N$_2$H$^+$ in second order by including 31 ionization channels for collision energies up to  1 eV. The rotational effects are neglected. The obtained cross sections for vibrationally relaxed target ($v^+_i=0$) are presented in fig. \ref{fig:1dxs}. The violet and magenta curves stand for the direct and total cross sections. Our results confirms that the direct mechanism is negligible and the indirect process is driving the DR for N$_2$H$^+$. The gain obtained by including the Rydberg resonances is more than two orders of magnitude. Although this finding is very important,  our total cross section underestimates the CRYRING measurements with more than two orders of magnitude, meaning that the 1D model with frozen NN bond is not suitable for quantitively describe the DR process. Moreover, besides the stretching mode one has to consider the bending mode of the linear molecule too. In conclusion, an accurate quantitative study needs a 3D description.

\section{Dissociative recombination of N$_2$H$^+$: The 3D model}\label{sec:3D}

The multi-dimensional nature of the vibrational motion of polyatomic ions makes the theoretical study of their dissociative recombination more difficult than that of the diatomic. 

We explore the complex dynamics of the DR in this case with a method based on the normal mode approximation combined with the R-matrix theory~\cite{tennyson2010} and vibronic frame transformation within MQDT. The normal mode approximation is responsible for the vibrational manifold, and usually provides a good description of the vibrational dynamics of molecular systems near the equilibrium geometry. 
We perform the R-matrix calculations in order to describe the electron scattering on the target. This provides us the reaction matrix, avoiding the {\it a priori} production of the PECS of the  electronic states of the neutral and their electronic couplings with the ionization continuum. Eventually, we perform the frame transformation from the body-fixed (or molecular) frame to the laboratory frame.

\begin{table}[t]
	\begin{center}
	\caption{
Equilibrium geometries, normal mode frequencies and  and permanent dipole moments of the linear N$_2$H$^+$ cation calculated with the MOLPRO quantum chemistry suite. Our results are compared with other calculations and with experimental result. $\ddagger$ stand for a CASSCF-MRCI calculation. 
	}
	\label{tab:1}
	\begin{tabular}{llcccccc}
		\hline
		\hline
		basis & theory & $R_{\mbox{NN}}$ & $R_{\mbox{NH}}$ & $\omega_{1,2}$  & $\omega_{3}$ & $\omega_{4}$ & $\mu_e$\\
		         &           & ($\AA$) & ($\AA$) & (cm$^{-1}$) & (cm$^{-1}$)  & (cm$^{-1}$)  & (debye)\\
		\hline
		\hline
		    & HF &  1.0606 & 1.0218 & 823.80 & 2649.61 & 3594.16 & 3.360\\
		    & CCSD(T) & 1.0992 & 1.0343 & 701.86 & 2280.39 & 3409.71 & 0 \\
		    & MRCI &  1.0980 & 1.0312 & 714.44 & 2285.17 & 3433.82 & 3.361 \\
		    & MRCI$^\ddagger$ & 1.0998 & 1.0337 & 701.78 & 2273.58 & 3412.08 & 3.365 \\
cc-pvtz	    & CEPA(2) & 1.0981 & 1.0338 & 688.80 & 2282.65 & 3411.65 & 3.470 \\
		    & MP2 &  1.1082 & 1.0349 & 704.49 & 2146.96 & 3393.49 & 3.487 \\
		    & MP4 & 1.1091 & 1.0348 & 680.15 & 2126.38 & 3393.11 & 0 \\
		    & CISD & 1.0835 & 1.0282 & 752.03 & 2443.90 & 3492.74 & --- \\
		    & CASSCF & 1.1005 & 1.0395 & 710.58 & 2274.15 & 3368.87 & 3.419 \\
		    \hline
theory~\cite{Hickman2011,Kashinski2012}  & 1D model & 1.120 & 1.074 & --- & --- & --- & ---\\
theory~\cite{Santos2014}  & CAS-Cl & --- & --- & 641 & 2252 & 3383 & 3.381 \\
exp.~\cite{kaddadj1994} & & --- & --- & 693 & 2269 & 3400 & ---\\
exp.~\cite{green1974} & & --- & --- & --- & --- & --- & 3.40\\
		\hline
		\hline
	\end{tabular}
	\end{center}
\end{table}

In our calculation we follow the general ideas presented in \cite{Santos2014}, relying on the following assumptions: (1) the rotation of the molecule is neglected, (2) the cross section is averaged over the autoionizing resonances, (3) the autoionization lifetime is assumed to be much longer than the predissociation lifetime and (4) the harmonic approximation is used to describe the vibrational states of the core ion. The electron induced recombination is described in the same way as its excitation but, instead of leaving the vibrationally excited ion, the electron is captured into a Rydberg resonance attached to the vibrational state excited by the electron. If the electron is captured by the ion, the system will most likely dissociate, rather than autoionize. In other words, if the energy of the electron is not sufficient to excite the ion and leave it, the probability of excitation is identical to that of dissociation. Correspondingly, the DR cross section reads as follows:
\begin{equation}
\label{eq:DRxsec3}
\sigma^{DR}(\varepsilon)=\frac{\pi}{4  \varepsilon}\sum_{i=1}^{n_{modes}} \theta(\hbar\omega_i-\varepsilon)g_i\sum_{ll'\lambda\lambda' }\left\vert \frac{\partial S_{l\lambda,l'\lambda'}}{\partial {q}_i}\right\vert^2\,.
 \end{equation}
Here $\theta$ is the Heaviside step function, $i$ runs over all the modes of the target molecule, $g_i$ is the degeneracy factor for the mode $i$, 
$l,\lambda$ and $l',\lambda'$ are the indices of the partial waves and their projections on a chosen quantization axis in the molecular frame of reference. The above formula gives the DR cross section assuming that the target ion is on its ground vibrational level. More details about the theory can be found for example in \cite{slava2018,mezei2019}. 

The 14 electrons of the linear triatomic N$_2$H$^+$ in the equilibrium geometry of the ground electronic $^1\Sigma$ state are distributed in the electronic configuration $(1\sigma)^2(2\sigma)^2(3\sigma)^2(4\sigma)^2(5\sigma)^2(1\pi)^4$. It has three normal modes ($n_{modes}=3$): the symmetric ($q_4,\omega_4$) and asymmetric stretching ($q_3,\omega_3$) and the doubly degenerate bending ($q_{1,2},\omega_{1,2}$). Table~\ref{tab:1} contains the relevant output of the electron structure calculations we have performed on the target by using the Molpro program suite~\cite{Werner2011} in the framework of the normal mode approximation. Using cc-pVTZ basis functions at different levels of theory, we have calculated the ground state energy in the optimized equilibrium geometry, the normal mode frequencies and the permanent dipole moments. 

\begin{figure}[t]
\centering
\includegraphics[width=0.8\linewidth]{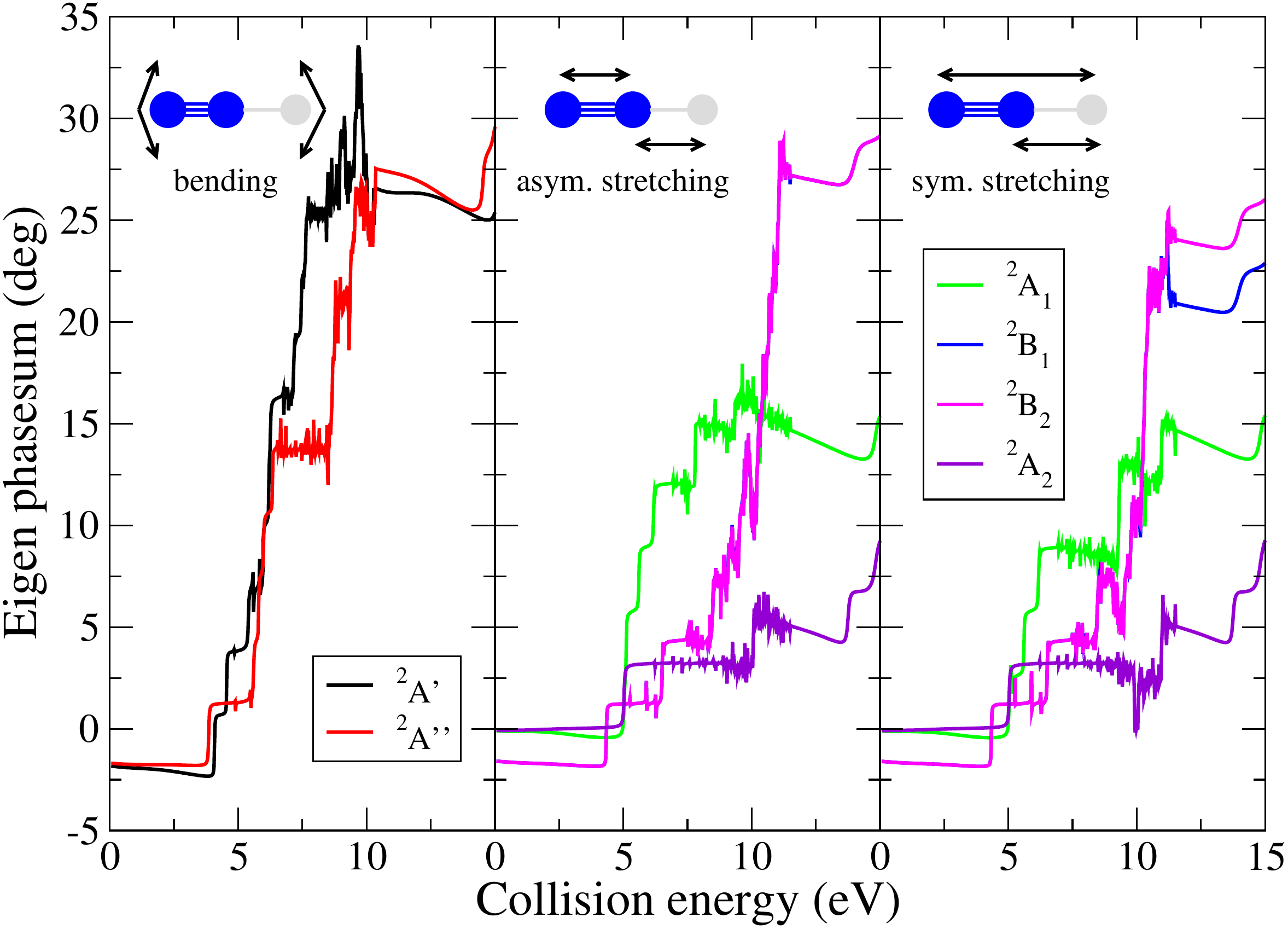}
 \caption{ The sum of eigenphases for the three different normal modes as a function of the electron scattering energy for a displacement of $q_i=0.05$ along each normal mode.
 }
\label{fig:7}
\end{figure}

The characterization of the target is followed by electron scattering calculations performed on the "electron+target" system. This part is the bottleneck of our method, since most of the higher level theories and bases functions are not tractable for the scattering calculations performed with UK R-matrix based Quantemol-N programme suite~\cite{Carr2012,Tennyson2007}. The calculations were performed in the abelian subgroup $C_s$ (bending mode) and $C_{2v}$ (stretching modes) and the target ion was assumed to be in its ground electronic state. We have performed Configuration Interaction Self-Consistent Field calculations using Hartree-Fock orbitals by freezing 6 electrons in the core $1(a')^2, 2(a')^2, 3(a')^2$ molecular orbitals (MOs) for bending mode, and $1(a_1)^2, 2(a_1)^2, 3(a_1)^2$ for the stretching modes. The remaining 8 electrons  are kept free in the active space of the $4a', 5a', 6a', 7a', 1a'', 2a''$  MOs for bending and $4a_1, 5a_1, 6a_1, 1b_1, 2b_1, 1b_2, 2b_2$ MOs for stretching modes, respectively. Virtual molecular orbitals have been added to the Complete Active Space for the augmentation to the continuum orbitals in the following way: 3 virtual MOs of $A'$ and 1 virtual MO of $A''$ symmetries for the bending mode and 3 $A_1$, 1 $B_1$ and 1 $B_2$ virtual MOs for the stretching ones. A total of nine electronic excited target states are included in our close coupling calculation. We used an R-matrix sphere of radius 14 bohr and in the partial wave expansion we went up to $l=4$. 

\begin{figure}[t]
\centering
\includegraphics[width=0.8\linewidth]{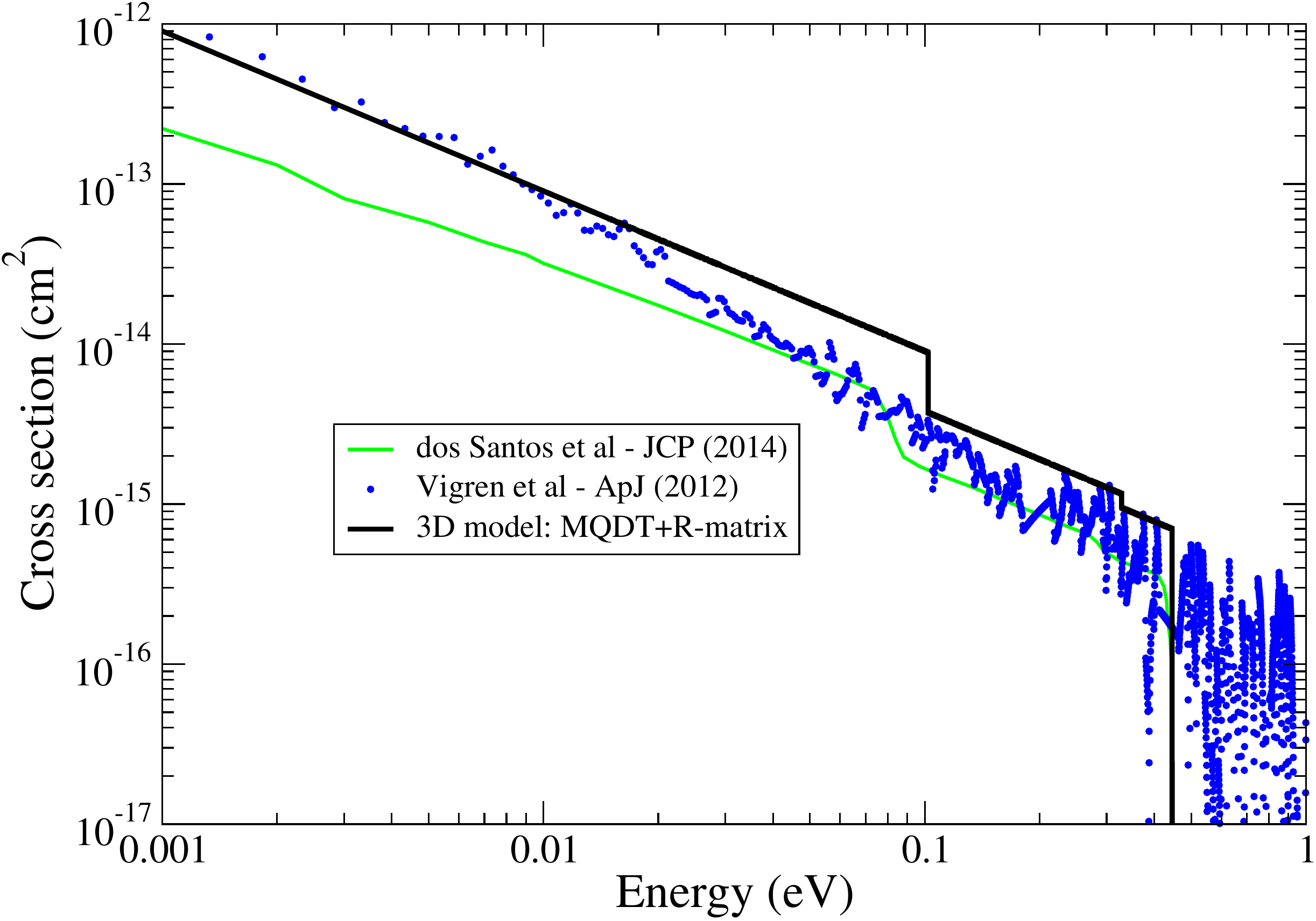}
 \caption{DR of N$_2$H$^+$ at low energy of the  incident electron within the 3D model. Blue symbols: CRYRING measurements~\cite{Vigren2012}.  Green line: results obtained with the complex Kohn variational method combined with the normal mode approximation~\cite{Santos2014}. Black line: present results relying on R-Matrix - normal mode approximation combined with MQDT frame transformation.
 }
\label{fig:8}
\end{figure}

At low collision energies the fixed-nuclei scattering matrix depends only weakly on energy. A quantity convenient for the analysis of the energy dependence of the scattering matrix is the eigenphase sum. Figure~\ref{fig:7} displays the eigenphase sum for the three different normal modes corresponding to a small displacement from equilibrium geometry ($q_i=0.05$) along each normal mode. The variation of the eigenphase sums is smooth for energies below 4.5 eV, suggesting the importance of the indirect mechanism.

The elements of the scattering matrix at a given geometry are computed from the reactance matrix by performing a Cayley transform, and they are expanded to first order in the normal coordinates~\cite{slava2018}. To calculate the derivative of the scattering matrix with respect to the normal coordinate, the scattering matrix is evaluated for two coordinate values of each normal mode, namely for  $q_i=0.05$ and $0.5$.

Figure~\ref{fig:8} shows in black the calculated DR cross sections of N$_2$H$^+$ in the framework of the 3D model, by applying the formula~(\ref{eq:DRxsec3}). 
Our cross section is compared with the CRYRING~\cite{Vigren2012} measured results (blue dots) and with that theoretically produced by Santos {\it et al} (green curve)~\cite{Santos2014}. 

At very low scattering energies, i.e. below 30 meV, the DR cross section is a smooth function inversely proportional to the incident energy of the electron, as predicted by the Wigner threshold law, whereas at higher energies, it exhibits a drop at each vibrational threshold. Our cross section is in very good agreement with the experimental results at low collision energy, below 30 meV, while at higher collision energies the agreement still remains within a factor of two. Moreover, our calculated cross section shows similar features as those obtained with the complex Kohn-variational method based 3D model~\cite{Santos2014}.

\section{Conclusion}

In the present work we have studied the dissociative recombination of the astrochemically relevant diazenylium, N$_2$H$^+$ molecular cation. 

By using a 1D model with frozen NN bonds we have calculated DR cross sections in the framework of the Multichannel Quantum Defect Theory and pointed out the importance of the indirect mechanism for this process. The obtained total cross section underestimates the storage-ring measurements with more than two orders of magnitude, implying that the 1D model is not sufficient for quantitively describe the DR process. Moreover, besides the stretching modes one has to consider the bending mode of the linear molecule. 

Consequently, we have performed a 3D model calculation in the framework of the normal mode approximation combined with R-matrix theory and Multichannel Quantum Defect Theory. Our results including all three normal modes agree well with the storage-ring measurements and, at very low energy, improve the results of the previous theoretical calculations, relevant for the interstellar media kinetics. The present study reveals the importance of treating the N$_2$H$^+$ molecule in its full dimensionality (3D model).

A major issue in the present astrochemical studies is the magnitude of the isotopic effects in the N$_2$H$^+$ DR, with respect to both isotopes of N and of H. Calculations on this issue are the subject of an ongoing project.

\begin{acknowledgments}
This study originates in the very stimulating scientific environment of the French consortium GdR THEMS.
The authors acknowledge support from Fédération de Recherche Fusion par Confinement Magn\'etique (CNRS and CEA), La R\'egion Normandie, FEDER, and LabEx EMC3 via the projects PTOLEMEE, Bioengine COMUE Normandie Universit\'e, the Institute for Energy, Propulsion and Environment (FR-IEPE), the European Union via COST (European Cooperation in Science and Technology) actions TUMIEE (CA17126), MW-Gaia (CA18104) and MD-GAS (CA18212).
The authors are indebted to Agence Nationale de la Recherche (ANR) via the project MONA. This work was supported by the Programme National “Physique et Chimie du Milieu Interstellaire” (PCMI) of CNRS/INSU with INC/INP co-funded by CEA and CNES. 
MAA and VK acknowledge support from the Thomas Jefferson Fund of the Office for Science and Technology of the Embassy of France in the United States and the National Science Foundation, Grant No. PHY-2102188 and the program ``Accueil des chercheurs \'{e}trangers'' of  CentraleSup\'{e}lec.
JZsM. and OA thank the financial support of the National Research, Development and Innovation Fund of Hungary, under the K18 and FK19 funding schemes with project no. K 128621 and FK 132989. 
JZsM, OA and IFS are grateful for the support of the NKFIH-2019-2.1.11- TÉT-2020-00100 and Campus France-Programme Hubert Curien+BALATON+46909PM Projects.
JZsM is  grateful for IEPE for a Visiting Professor Fellowship.
\end{acknowledgments}

\section*{Data Availability}
The data underlying this article will be shared on reasonable request to the corresponding author.

%\nocite{*}
%\section*{References}
%\bibliography{Schneider-N2-4arxiv.bib}% Produces the bibliography via BibTeX.

\end{document}